\magnification=\magstep1
\voffset=0.5 true cm
\hoffset=0.66cm
\hsize=15.0 true cm
\baselineskip=12pt
\parindent=13pt
\headline={\ifnum\pageno=1\hss\else\hss - \folio\ - \hss\fi}
\footline={\hss}
\font\meinfont=cmbx10 scaled \magstep1  
\noindent {\meinfont Einstein-Podolsky-Rosen Correlations in  Deu\-ter\-on
Photodisintegration}
\par
\medskip\bigskip
\noindent  Arthur Jabs
\medskip
\par
\noindent  Physics Department, Federal University of  Para\'\i ba, Jo\~ao 
Pessoa, 
Brazil
\par
\medskip
\noindent
(April 1996)
\par
\bigskip\medskip
\leftskip=1.0 true cm
\noindent
{\bf Abstract.} Einstein-Podolsky-Rosen  correlations  have  so  far   been
measured only between  pairs  of  photons  and  between  pairs  of
protons. It is proposed to measure these correlations between  the
proton and the neutron emerging from the breakup of  the  deuteron
induced by gamma rays near threshold. The feasibility of the experiment
is discussed. Polarimeters with substantially higher overall efficiency
 than the presently reported value of about $10^{-4}$ are needed in order to
get enough events.

\par
\medskip
\noindent  PACS:
25.85.Jg; 03.65.Bz
\medskip\bigskip
\par
	
\leftskip=0pt
\noindent {\bf 1~~Introduction} \smallskip\par
\noindent The interesting feature of the Einstein-Podolsky-Rosen (EPR) correlations between spatially
separated particles is that  they  imply  a  violation  of  Bell's
inequality and hence reveal a certain fundamental  nonlocality  of
nature [1]. This is predicted by quantum mechanics and has by  now
been  confirmed  in  quite   a   number   of   expriments   [2].
Nevertheless, all those experiments are only concerned with  pairs
of photons, except one, which is concerned with pairs  of  protons
[3]. In  view  of  the  importance   of   nonlocality   for   our
understanding of nature and in view  of  the  fact  that  specific
shortcomings  of  the   individual   experiments   still  permit  loopholes  to  escape  from  the
acceptance of nonlocality [4], it is desirable  to  investigate  a
larger variety of physical situations. Besides this I have a special reason for suggesting this experiment because I have proposed an alternative to the Copenhagen interpretation in quantum mechanics in which I predict that no EPR correlations will show up between elementary particles that are not identical and not particle-antiparticle pairs [5]. Be that as it may, the experiment will be
interesting in  itself.
\par
A number of proposals which involve  particles other than photons and protons 
have already been made, though not yet  performed.
 The N  and  the O   atom   obtained   by
(pre)dissociation of an NO molecule have been considered in [6], and the Na  atom  pairs  obained  by  dissociation  of Na$_{2}
$ molecules in [7]. And there exists by now quite a number of more or  less
detailed proposals concerned with particle-antiparticle pairs such
as e$^{+}$e$^{-}, \mu ^{+}\mu ^{-}, \tau^{+}\tau^{-},{\rm K}\overline{\rm K}, \Lambda \overline{\Lambda }$ and B$^{\rm o} \overline{{\rm B}}^{\rm o}$ [2,8,9].
\par
The proposed experiment  differs  from  the
above-mentioned ones in that it is concerned (1) not  with  atoms  but
with  elementary  particles,  and (2)  with elementary particles   that   are   not
particle-antiparticle pairs.
\par\medskip\bigskip
\vfill\eject
\noindent
{\bf 2~~Typical Experimental Situation}
\smallskip\par\noindent
A  typical  experimental  arrangement   for   measuring   EPR
correlations are two  spin-${1\over 2}$  particles  which  originate  from  a
collision or decay in a  spin  singlet  state  and  fly  apart  in
opposite directions. Particle 1 then enters the Stern-Gerlach-type
apparatus  A  and particle 2 apparatus B. Particle 1 is  deflected
either upward or downward with respect to the spin-reference  axis {\bf a} of apparatus A and shows up with either spin up $(r_{\rm A}=+1)$ or  spin
down $(r_{\rm A}=-1)$. The same applies to particle 2 in apparatus B  with
axis {\bf b}. {\bf a} and {\bf b} are unit vectors. If $E({\bf a}, {\bf b})$ 
is the average of the product $r_{\rm A}r_{\rm B}$  the  Bell
inequality (in one of its many equivalent forms) reads [1]
$$
\vert E({\bf a},{\bf b}) + E({\bf a},{\bf b}') + E({\bf a}',{\bf b}) - E({\bf a}',{\bf b}')\vert  \le  2 .
\eqno(1)$$
\noindent Let $P(r_{\rm A},r_{\rm B}\vert  \vartheta )$ be the probability of particle 1 being pushed  into
a spin-$r_{\rm A}$ state and particle 2 into a spin-$r_{\rm B}$ state,  where $\vartheta\; (0 \le \vartheta \le \pi) $  is
the angle between the axes {\bf {a}} and  {\bf {b}}.  Standard  quantum  mechanics
predicts the expressions
$$
P(r_{\rm A},r_{\rm B}\vert  \vartheta ) = \hbox{${1\over 4}$}(1 - r_{\rm A}r_{\rm B}\cos \vartheta ) ,
\eqno(2)$$
$$
E({\bf a},{\bf b}) = - {\bf ab} = - \cos \vartheta
\eqno(3)$$
\noindent for the singlet state [1]. This can lead to a violation  of  Bell's  inequality for certain
choices of $\vartheta $ and is therefore locally inexplicable [1].
\medskip\bigskip
\par\noindent
{\bf 3~~The Deuteron Case. Singlet State}
\smallskip\par\noindent
The above  situation  can  easily  be  arranged  in  the
photodisintegration of the deuteron induced  by  photons  with LS 
(laboratory system) energies $E_{\gamma }$ above the threshold of 2.226  MeV.
Up to 2.3 MeV the total cross section rises as [10,11]

\midinsert\vskip 9 true cm
\par\noindent
Fig.~1. Deuteron photodisintegration cross section between threshold and 18
MeV  photon energy. From [12]
\bigskip\endinsert

$$
\sigma = 675 {\sqrt{E_\gamma/E_{\rm u} - 2.226} \over (E_\gamma/E_{\rm u})\, (E_\gamma/E_{\rm u} - 2.149) } \hbox{~~mb}, \qquad \qquad E_{\rm u}= 1\, {\rm MeV}.
\eqno(4)$$
The cross section reaches its maximum of 2.5 mb at about 4.4 MeV, and at 22 MeV it has decreased to 0.5 mb. Up to $\approx$ 2.4 MeV formation of the spin singlet state (M1 transition, $^3\!S\rightarrow ^1\!\!S$, formula (4)) prevails, then that of the triplet state (E1 transition, $^3\!S \rightarrow ^3\!\!P$) [11,12].
\par
Let us first assume that only the singlet state contributes. The modification due to some contribution of the triplet state will be considered later. To measure  the spin correlation one may follow the procedure of the proton-proton scattering experiment [3]. In that experiment protons of 13 MeV kinetic LS energy  were scattered by a hydrogen target. Each of the two protons emerging from the scattering was slowed down to 6 MeV and then entered a spin-component analyzer (polarimeter). In this analyzer it was scattered by a carbon foil and then registered by one of two detectors (R or L). These detectors lay in a plane which contained the direction of the proton entering the analyzer. Each detector formed a fixed angle of $50^\circ$ with that direction, and the plane with the detectors could be rotated around that direction. The coincidences between the detectors of one analyzer with the detectors of the other were counted 
($ N_{\rm LL}, N_{\rm LR}$, etc.). The expression
$$
E_{\rm exp} = { N_{\rm LL} + N_{\rm RR} - N_{\rm RL} - N_{\rm LR} \over
N_{\rm LL} + N_{\rm RR} + N_{\rm RL} + N_{\rm LR} },
\eqno(5)$$ 
after some corrections, represents the experimental value for $E({\bf a}, {\bf b})$. It was measured for various angles $\vartheta$ between the detector planes of the two analyzers (in the center-of-momentum system (CMS) of the protons) and was compared with the maximum value $E_{\rm max}(\vartheta)$ compatible with Bell's inequality. $E_{\rm max}(\vartheta)$ was calculated for a selected number of angles, observing invariance of {\it E} under reflection and rotation. A smooth interpolation formula is [8]
$$
E_{\rm max}(\vartheta) = 2 \vartheta / \pi - 1.
\eqno(6)$$
This formula, by the way, coincides with the spin correlation formula calculated for the two fragments of an exploding classical bomb [13]. The maximal difference between $\vert E(\vartheta) \vert$ of the formulas (3) and (6) is 0.21 and occurs at $\vartheta = 40^\circ$. The experimental values confirmed the quantum mechanical formula (3) and for some values of $\vartheta$ were definitely outside the range (6) allowed by Bell's inequality. In the deuteron experiment the role of the two protons after scattering is taken over by the proton and neutron after disintegration and the same procedure of comparing $\vert E(\vartheta)\vert$ with $\vert E_{\rm max}(\vartheta) \vert$ may be adopted here.
\smallskip\bigskip\par\noindent
{\bf 4~~The Triplet Contribution}
\smallskip\par\noindent
If the nucleons emerging from deuteron disintegration are in a spin triplet state the spin correlation formulas are more complicated. Quantum mechanics in this case predicts [14]
$$
E({\bf a},{\bf b}) = a_z b_z \qquad \qquad \hbox{in~~} \vert 1,+1\rangle \hbox{~~and~~} \vert 1,-1 \rangle
\eqno(7)$$
$$
E({\bf a}, {\bf b}) = {\bf ab} - 2a_z b_z \; \hbox{~~in~~} \vert 1,0 \rangle \hbox{~~~~~~~~~~~~~~~~~~~~}
\eqno(8)$$
where $a_z$ and $b_z$ are the components of {\bf a} and {\bf b} in some preferred direction. This direction, that is, the exact wave function or density matrix of the triplet state must also be known, and this requires additional experimental efforts. One might therefore restrict oneself to situations where the nucleons emerge in a pure singlet state. As mentionad in the introduction this occurs when the $\gamma$-ray energy is only little above threshold. On the other hand, the cross section for disintegration goes down rapidly when $E_\gamma$ approaches the threshold, formula (4). One may therefore retain somewhat higher $\gamma$-ray energies, where the cross section is larger, but the contribution from the triplet state is still small. Observing that the angular distribution of the nucleons in the singlet state is isotropic and in the triplet state  follows a $\sin^2 \Theta$ law ($\Theta$ = angle between the proton and the direction of the $\gamma$ rays) one may restrict oneself to protons and neutrons em
\par
We want to consider this point more quantitatively.
An  estimate  of  the  relative  triplet  contribution  as  a
function of the  photon  energy $E_{\gamma }$  can  be  extracted  from  the
empirical and theoretical data presented in [11,12,15,16]: in  the
range of $E_{\gamma }= 2.3-5$ MeV the CMS differential cross  section can be written as
$$
d\sigma /d\Omega  = a_{\rm M} +a_{\rm E} + (b_{\rm M} +b_{\rm E})\sin ^{2}\Theta 
\eqno(9)$$
\noindent where  the  index M  denotes  the  singlet  and E  the   triplet
contribution, and
\par
$$
a_{\rm M}/a_{\rm E} = 48\times (E_{\gamma }/E_{\rm u} - 2.226)^{-1.75}
\eqno(10)$$
$$
b_{\rm E}/a_{\rm E} = 6910\times (E_{\gamma }/E_{\rm u})^{-1.89}
\eqno(11)$$
$$
b_{\rm M} = 0
\eqno(12)$$
$$E_{\rm u} = 1 \hbox{~MeV}$$
\noindent represent simple numerical fits.  From  this  one
obtains, for example, that the relative  triplet  contribution  in
forward direction $\Theta  = 0^{\circ}$
$$
(d\sigma /d\Omega )_{\rm E}/[ (d\sigma /d\Omega )_{\rm E} +(d\sigma /d\Omega )_{\rm M} ] = (1 + a_{\rm M}/a_{\rm E})^{-1}
\eqno(13)$$
\noindent at $E_{\gamma } =  5$ MeV reaches $11\%$.
\par
When a finite opening (half) angle $\alpha $ around $\Theta  = 0^{\circ}$  is  taken
into account the relative triplet contribution becomes larger. One
has  to  replace \ $d\sigma /d\Omega $ \   by\  $\sigma /\Omega 
   =   \int ^{\alpha }_{o}\int ^{2\pi }_{o}
(d\sigma /d\Omega )\sin \Theta d\Theta d\varphi  /
 \int ^{\alpha }_{o} \int ^{2\pi }_{o}\sin \Theta \* d\Theta \* d\varphi $. 
Using (9) one obtains
$$
\sigma /\Omega  = a_{\rm M}+a_{\rm E} +(b_{\rm M}+b_{\rm E})\times (\hbox{${2\over3}$} +\hbox{${1\over 3}$}\cos ^{3}\alpha  -\cos \alpha )/(1-\cos \alpha )
\eqno(14)$$
$$
\approx  a_{\rm M} +a_{\rm E} + \hbox{${1\over 2}$} (b_{\rm M} +b_{\rm E})\alpha ^{2}
\eqno(15)$$
\noindent where the error of the approximation (15) is less than $5\%$ up to $\alpha  =
25^{\circ}$. From (15) and (12) one obtains now the relative triplet contribution at an opening angle $\alpha$ around the forward direction
$$
(\sigma /\Omega )_{\rm E}/[(\sigma /\Omega )_{\rm E} 
+ (\sigma /\Omega )_{\rm M}] = 
\left(1 + {a_{\rm M}\over a_{\rm E}} 
{1\over 1+(b_{\rm E}/a_{\rm E})(\alpha ^{2}/2)}\right)^{-1}.
\eqno(16)$$

\noindent Using (10) and (11) and chosing an opening angle of $\alpha  = 2^{\circ}$  one
obtains a contribution of  $13\%$  at $E_\gamma = 5$  MeV,  which is 
not much larger than the contribution of $11\%$  obtained
for $\alpha  = 0^{\circ}$ from (13).
\par
Now, a contribution of less than 18\% of the triplet state is still acceptable because one usually expects the correlations to be in accordance with quantum theory, which in this case predicts values for $\vert E({\bf a},{\bf b}) \vert$ that are still larger than $\vert E_{\rm max}({\bf a},{\bf b})\vert$ for some angles $\vartheta$. This is seen in the following way: the inclusion of the triplet state may lead to a smaller value of $\vert E({\bf a},{\bf b}) \vert$ that that calculated for the singlet state only. Let us write (in obvious notation)
$$
E_{\rm st}({\bf a},{\bf b}) = (1 - r) E_{\rm s}({\bf a},{\bf b}) + 
r E_{\rm t}({\bf a},{\bf b})$$
and let us assume the extreme case that $E_{\rm t}({\bf a},{\bf b})$ is just the negative of $E_{\rm s}({\bf a},{\bf b})$ (cf. formula (8) with $a_zb_z=0$). \ 
 {\it r} is the relative overall contribution of the triplet state. With formula (3) this gives us $E_{\rm st}({\bf a},{\bf b}) = E_{\rm st}(\vartheta)=(2r-1)\cos \vartheta$, and with $E_{\rm max}(\vartheta)$ from formula (6) we obtain the difference
$$
\Delta = \vert E_{\rm st}(\vartheta) \vert - \vert E_{\rm max}(\vartheta) \vert = (1-2r)\cos \vartheta + 2\vartheta/\pi -1
\eqno(17)$$
where the last expression holds for $r \le 0.5,\, \vartheta \le 90^{\circ}$. For $r=0$ (singlet only) formula (17) results in $\Delta = 0.21 \;  (\vartheta=40^{\circ})$. For larger $r \; \Delta$ decreases. It reaches zero at $r=0.18 \; (\vartheta=90^{\circ})$, but it is still $\Delta =0.07 \hbox{~for~} r=0.1 \; (\vartheta =50^{\circ})$. Thus, the better the experimental accuracy the smaller the value of $\Delta$ that can be resolved, and the larger the value of {\it r} that can still lead to a demonstration of a violation of Bell's limit. Of course, with this procedure, considering only an upper limit of the triplet contribution, we cannot demonstrate that the violation is just the one predicted by quantum mechanics. Nevertheless a violation of Bell's inequality can in principle be demonstrated.
\smallskip\bigskip\par
\vfill\eject
\noindent
{\bf 5~~The Rate of Events}
\smallskip\par\noindent
If the experiment is done in the way discussed above, comparing $\vert E(\vartheta)\vert \;\allowbreak  \hbox{with} \allowbreak \; \vert E_{\rm max}(\vartheta) \vert$, the accuracy is determined by that of $E_{\rm exp}$ from formula (5). The $N_{\rm LL}$ etc. are numbers of coincidences, and the higher these numbers the better the statistical accuracy. In the following we shall only consider this statistical accuracy, that is, the rate of coincidence events. This is always an important point in experiments of this nature.
\par
 The rate is proportional  to the product of the incoming $\gamma$-ray flux, the number of target deuterons, the photodisintegration cross section, and the square of the overall efficiency of the polarization analyzers. In order to obtain a rough estimate let as assume that the deuteron experiment is done in exactly the same way as the pp-scattering experiment [3], so that we only have to replace the flux of incoming protons ($5\times 10^{18}/{\rm m}^2 {\rm s}$) by that of incoming photons ($10^{18}$/MeVs, [17], assuming a 1 MeV $\gamma$-ray energy range and assuming that all of the beam's cross section can be utilized) and the pp cross section ($5\times10^{-29}\, {\rm m}^2$) by the photodisintegration cross section ($2.5\times 10^{-31}\, {\rm m}^2$, maximum value). The number of target particles is assumed to be the same in both experiments ($1.4\times 10^{19}$). These numbers result in coincidence rates in the deuteron experiment that are smaller by a factor $10^{10}$ than those in the pp experiment. The 
\par
The main reason for the low coincidence rates, both in the deuteron and in the pp experiment, is the low ``transmission'' or ``overall polarimeter (analyzer) efficiency''. This efficiency is defined as the number of particles that are deflected and registered in one of the detectors of the analyzer divided by the total number of particles that entered the analyzer. It is of the order of $10^{-4}$ for the considered protons and neutrons of a few MeV [18]. This value contributes quadratically because the number of coincidences between two equal analyzers is proportional to the square of their efficiency.
\par
Since the polarimeters utilize the scattering of nucleons in carbon foils their low efficiencies are mainly due to the small cross section for this process, which is of the order of a barn ($10^{-28}{\rm m}^2$) for nucleons of a few MeV. As far as I can see there might still be hope to obtain larger efficiencies when one considers nucleons of lower energy, down to 100 keV or even less. At these energies the nucleon-carbon cross sections increase considerably, especially for protons, and they exhibit strong fluctuations and large bumps at nuclear resonances. So one might perhaps find special energy values where these cross sections are large and lead to high efficiencies. Or one may be able to invent novel types of polarimeters for these slow nucleons. (Stern-Gerlach magnets for neutrons?). This, together with an effort to get a high flux of incoming $\gamma$ rays and a large number of target deuterons, might lead to sufficient events in a reasonable time.
\par
Another aspect of a low efficiency is that it permits a loophole to escape from accepting nonlocality [3]. Even if the efficiency is high enough to lead  to a sufficient number of events it may not be high enough to close that loophole. I think, however, that even so the experiment would be interesting enough.
\par
Slow nucleons from deuteron photodisintegration  means $\gamma$ rays near threshold, with the additional advantage of a pure singlet state. Here, in very good approximation it is $E_{\gamma}= 2.226 \, {\rm MeV} + 2 E_{\rm kin}$, where $E_{\rm kin}$ is the kinetic energy of one nucleon, hence $E_{\gamma} > 2.4$ MeV for $E_{\rm kin} > 100$ keV. Relativistic kinematics [19] shows that for 2.4 MeV $< E_{\gamma} < 35$ MeV the velocity of the nucleons in the CMS is always more than 10 times the velocity of the CMS in the LS. So, in the CMS$\leftrightarrow$LS transformation the angles never change by more than 10\% and the kinetic energies never by more than 20 \%.  

\par
\medskip
I thank Prof. Dr. G. S\"u\ss mann and the Physics Section  of  the
University of Munich for their hospitality.  My  interest  in  the
deuteron arose from a discussion with Prof. S\"u\ss mann and Dr.  B.-G.
Englert.
\par
\bigskip\smallskip

\noindent {\bf References}
\smallskip
\parindent=16pt 
\par\noindent\hangindent=16pt\hangafter=1
~1.~Bell, J. S.: Speakable and unspeakable in  quantum  mechanics,
p. 139. Cambridge: Cambridge University Press 1987
\par\noindent\hangindent=16pt\hangafter=1 
~2.~Home, D., Selleri, F.: Riv. Nuovo Cimento {\bf 14}, 1 (1991)
\par\noindent\hangindent=16pt\hangafter=1 
~3.~Lamehi-Rachti, M., Mittig, W.: Phys. Rev. D {\bf 14}, 2543 (1976)
\par\noindent\hangindent=16pt\hangafter=1
~4.~Selleri, F., Tarozzi, G.: Riv. Nuovo Cimento {\bf 4}, 1 (1981)
\par\noindent\hangindent=16pt\hangafter=1
~5.~Jabs, A.: Brit. J. Phil. Science {\bf 43}, 404 (1992); Physics Essays {\bf 9}, 36 (1996)
\par\noindent 
~6.~Livi, R.: Nuovo Cimento {\bf 48B}, 272 (1978)
\par\noindent\hangindent=16pt\hangafter=1
~7.~Lo, T. K., Shimony, A.: Phys. Rev. A{\bf 23}, 3003 (1981)
\par\noindent\hangindent=16pt\hangafter=1 
~8.~T\"ornqvist, N. A.: in: Quantum  Mechanics
Versus Local Realism  (Selleri, F., ed.) p.115. New York: Plenum Press 1988
\par\noindent\hangindent=16pt\hangafter=1 
~9.~Home, D.: in: Ref. [8] p. 133; Tixier, M.  H.  et  al.:  Phys.
Lett. B {\bf 212}, 523 (1988); Datta, A., Home,  D.:  Found.  Phys.
Lett. {\bf 4}, 165 (1991); Privitera, P.:  Phys.  Lett B {\bf 275}, 172
(1992); Abel, S. A., Dittmar,. M., Dreiner, H.: Phys. Lett. B
{\bf 280}, 304 (1992)
\par\noindent\hangindent=16pt\hangafter=1
10.~Blatt, J. M., Weisskopf, V. F.: Theoretical Nuclear Physics, p. 608. New York: Wiley 1952 and Springer-Verlag 1979 and Dover 1991; Dawydow, A. S.: Theorie des Atomkerns, p. 391. Berlin: VEB Deutscher Verlag der Wissenschaften 1963 
\par\noindent\hangindent=16pt\hangafter=1
11.~Arenh\"ovel,  H.,  Sanzone,  M.:  Photodisintegration   of   the
Deuteron:  A  Review  of  Theory  and  Experiment  (Few  Body
Systems, Suppl. 3), Sections 6 and 7. Wien: Springer-Verlag 1991
\par\noindent\hangindent=16pt\hangafter=1
12.~Marmier, P., Sheldon, E.: Physics of Nuclei and Particles, Vol. II, p. 902. New York: Academic Press 1970
\par\noindent 
13.~Peres, A.: Am. J. Phys. {\bf 46}, 745 (1978)
\par\noindent
14.~Barut, A. O., Bo\v{z}i\'{c}, M.: Nuovo Cimento {\bf 101B},
595 (1988)
\par\noindent\hangindent=16pt\hangafter=1 
15.~Schmitt, K. M., Wilhelm, P., Arenh\"ovel, H.:  Few-Body  Systems
{\bf 10}, 105 (1991), Figs. 1, 12, 14
\par\noindent\hangindent=16pt\hangafter=1 
16.~Rustgi, M. L., Pandey, L. J.: Phys. Rev. C{\bf 40}, 1581 (1989)
\par\noindent\hangindent=16pt\hangafter=1
17.~Knei\ss l, U.: Low energy photo- and electrofission, in: Proc. of the Fourth Course of the Int. School of Intermediate Energy in Nuclear Physics, San Miniato 1983  (Bergere, R., Costa, S., Schaerf, C., eds.) p. 220, especially Table 1. Singapore: World Scientific 1984
\par\noindent\hangindent=16pt\hangafter=1
18.~Clegg, T. B.: New Developments in Technology and Facilities, in: Proc. Sixth Int. Symp. Polar. Phenom. in Nucl. Phys., Osaka 1985 (J. Phys. Soc. Jpn. {\bf 55}, 535 (1986))
\par\noindent\hangindent=16pt\hangafter=1 
19.~Baldin, A. M., Goldanskij, W. I., Rosental, I.  L.:  Kinematik
der Kernreaktionen. \S \S \  4, 5, 8. Berlin: Akademie-Verlag 1963

\bigskip
\centerline{\vrule width 5cm height0.4pt}
\bigskip
\bye